\documentclass[prl,reprint,flushbottom]{revtex4-1}
\usepackage{epsfig}
\usepackage{amsmath}
\usepackage{amssymb}
\usepackage{amsthm} 
\usepackage{bm}
\usepackage{color}
\newcommand{\col}[1]{\textcolor{red}{#1}}

\newcommand{\ket}[1]{\left|#1\right>}
\newcommand{\bra}[1]{\left<#1\right|}
\newcommand{\braket}[2]{\left< #1 \vphantom{#2} \right|\left. \!#2 \vphantom{#1} \right>}

\newcommand{\bea}{\begin{eqnarray}}
\newcommand{\ea}{\end{eqnarray}}
\newcommand{\eea}{\end{eqnarray}}
\newcommand{\ord}{{\cal O}}
\newcommand{\sumint}[1]

\begin{document}

\title{{``Photonic'' 
cat states from strongly 
interacting 
matter waves}}
\author{Uwe R. Fischer and Myung-Kyun Kang}

\affiliation{Seoul National University,   Department of Physics and Astronomy \\  Center for Theoretical Physics, 
151-747 Seoul, Korea}

\begin{abstract}
We consider ultracold quantum gases of scalar bosons,  residing in 
a coupling strength--density regime in which they constitute 
a twofold fragmented condensate trapped in a single well. 
It is shown that the corresponding quantum states are, in the appropriate Fock space basis, identical to the photon cat states familiar in quantum optics, 
which correspond to superpositions of coherent states of the light field with a phase difference of $\pi$. 
In marked distinction to photon cat states, the very existence 
of matter wave cat states however crucially depends on the 
many-body correlations of the constituent particles.
We consequently establish that the quadratures of the effective ``photons,"
expressing the highly nonclassical nature of the macroscopic 
matter wave superposition state,  
can be experimentally accessed by measuring the density-density correlations of the interacting quantum gas.
\end{abstract}

\pacs{
}

\maketitle

The eponymous  cat quantum states have originally been constructed 
by Erwin Schr\"odinger
to stimulate deeper thought on the sometimes surreal aspects of macroscopic quantum mechanics \cite{Schroedinger}.    
Present-day technology allows for going beyond the pure gedankenexperiment of Schr\"odinger's day to 
create, within the realm of quantum optics, small cat states,  Schr\"odinger ``kittens" \cite{OurScience}. 
They consist typically either in the superposition of coherent states of light \cite{Glauber} 
with a phase difference of $\pi$, coined photon cat states, 
or in states  $\ket{N0}+\ket{0N}$ of the NOON type \cite{Dowling}. 
Highly entangled photonic cat states 
constitute a possible basic building block  in quantum information architectures \cite{OurScience,Raimond,Ourjoumtsev,Afek,Gao,Yao}.

We propose in what follows a novel species of 
Schr\"odinger cat states, which are in distinction to photonic cat states 
relying on quantum many-body correlations, and  hence 
owe their very existence to strong interactions of the constituent bosons. 
They are generated in a scalar (one component)  gas 
of massive bosons trapped inside a single (harmonic or box) trap. 
The many-body states required to produce the cat states correspond to twofold
fragmented condensates, for which the single-particle 
density matrix has 
two $\ord{(N)}$ eigenvalues \cite{Penrose}. 
Condensate fragmentation \cite{Mueller} is by now 
firmly established as a many-body effect not describable within a mean-field theory
of the classic Bogoliubov type \cite{Alon,Alon2}. Specifically, for sufficiently large and positive 
contact interaction coupling, it derives from the broken translational symmetry and localization  
in a single trap \cite{Bader}. 
Therefore, strongly interacting matter wave cat states (termed SIMCAS in what follows) 
can be experimentally accessed, starting from a Bose-Einstein condensate (BEC),
 by  tuning the coupling with a Feshbach resonance \cite{Chin}.

Below, we shall demonstrate that SIMCAS, in the appropriate Fock space basis,  
form many-body states essentially indistinguishable from photonic cat states, for sufficiently large (but still mesoscopic) number $N$ of gas particles. 
The degree of fragmentation of condensate coherence 
will be shown to be directly related to the degree of 
macroscopicity of the coherent superposition of 
many-body states, quantified 
by the superposition size of the quantum 
state \cite{Leggett,Nimmrichter,Duer,Volkoff}. 
By evaluating quadratures \cite{Raimond}, we show that the 
macroscopicity of the matter wave quantum superposition is directly measurable through density-density correlations after time-of-flight expansion (TOF).

We emphasize that SIMCAS live in a single trap and not in a double well, 
and are obtained for a scalar, not a two-component gas, with {\em repulsive} interactions,  
which distinguishes them from previously suggested cat state implementations in dilute ultracold quantum gases, cf.   
\cite{Cirac,Dalvit,Dunningham,HoFrag,Huang}. 
Furthermore, 
SIMCAS  
are the 
{\em ground states} 
of the trapped 
Bose gas for relatively moderate interaction couplings, at which the single BEC crosses over to a two-fold fragmented condensate \cite{AlonPRL}  
(the gas remaining sufficiently dilute to ensure that three-body recombination rates remain small).
This renders SIMCAS different from the collapse-and-revival type quantum optics proposals to generate 
cat states 
in a Kerr medium cf., e.g., \cite{YurkeStoler,Kirchmair}, or involving 
scattering of matter waves at barriers \cite{Weiss}.

As a first step, we expand two-mode fragmented states in the Fock space basis of elementary bosons, 
\bea
\ket{\Psi}=\sum_{l=0}^NC_l\ket{N-l,l}, \qquad\sum_{l=0}^N|C_l|^2=1,
\label{origPsi}
\ea 
where the Fock space basis vectors $\ket{N-l,l}=\frac{(\hat{a}^{\dag}_0)^{N-l}(\hat{a}^{\dag}_1)^{l}}{\sqrt{(N-l)!l!}}\ket{{\rm vac}}$, with $\ket{\rm vac}$ the particle vacuum. 
{A two-mode Hamiltonian, with}   
interaction part $H_{\rm int} = \frac{A_1}2 \hat a^\dagger_0\hat a^\dagger_0 \hat a_0 \hat a_0 
+\frac{A_2}2 \hat a^\dagger_1\hat a^\dagger_1\hat a_1\hat a_1 +\frac{A_3}{2}\left(\hat a_0^\dagger\hat a^\dagger_0
\hat a_1\hat a_1+ {\rm h.c.}\right)  +\frac{A_4}2 \hat a_1^\dagger \hat a_1 
\hat a_0^\dagger\hat a_0$ (assuming that $A_3\in \Re$), 
valid, e.g., for contact and dipolar 
two-body interactions when the two modes have even and odd parity, respectively,   
generically leads to a distribution of the $C_l$ modulus which is, in the continuum limit, 
a Gaussian (of relative width $\propto 1/\sqrt N$), yielding fragmented condensate states for $A_3>0$ and 
$A_1+A_2+2A_3-A_4>0$ {(see detailed discussion in \cite{Bader})}.   
The {\em distribution center} $l_0:= \langle \hat a_1^\dagger \hat a_1 \rangle \in[0,N]$ quantifies the degree of fragmentation $\mathcal F=1-|\lambda_0-\lambda_1|/N$, where
$\lambda_i$ are eigenvalues of the single-particle density matrix.
We note that the \col{exact} functional form of the $C_l$ distribution is not important for the following argument. 
The conditions are (a) $0\ll l_0\ll N$, (b) negligible weight of $C_l$ at the boundaries $l=0,N$, 
and (c) small relative distribution width. 
The fragmentation degree is ${\mathcal F}=2l_0/N$  when $l_0\le N/2$ (assumed below without loss of generality) 
and $\mathcal F= 2(1-l_0/N)$ when $N/2 \le l_0\le N$ \cite{Bader,Kang}.

Self-consistent solutions of the quantum many-body problem have indeed also found such two-mode fragmented states 
for specific Hamiltonians, e.g.\,for quasi-one-dimensional (quasi-1D)  
BECs at sufficiently large contact interaction couplings \cite{AlonPRL,Chatterjee}.
The large overlap of the modes in a single trap, crucially, leads to maroscopically 
large interaction-induced pair-exchange processes, $-\big<\hat a_0^\dag \hat a^\dag_0 \hat a_1\hat a_1+{\rm h.c.} \big>\sim \ord(N^2)$ 
which {\em stabilize} (rather than destabilize) SIMCAS, as shown in \cite{Xiong}.   
The existence of negative macroscopic pair coherence due to a pair-exchange coupling $A_3\sim \ord (A_1,A_2,A_4)$ 
is thus a distinguishing feature of SIMCAS. 

To proceed, we construct the two {ladder operators} 
\bea
\hat{b}= \lim_{\epsilon \rightarrow 0} \frac{1}{\sqrt{\hat{N}_0+\epsilon}}\hat{a}^{\dag}_0\hat{a}_1, \quad
\hat{b}'=\lim_{\epsilon \rightarrow 0} \frac{1}{\sqrt{\hat{N}_1+\epsilon}}\hat{a}^{\dag}_1\hat{a}_0,
\label{def b}
\ea
where $\hat N_i = \hat a_i^\dag \hat a_i$.
The ladder operators and their Hermitian conjugates $\hat b^\dag,\hat b'^\dag$
convert particles between the two macroscopically occupied modes, according to 
$\hat{b}\ket{N-l,l}=\sqrt{l}\ket{N-l+1,l-1}, \quad \hat{b}^{\dag}\ket{N-l,l}=\sqrt{l+1}\ket{N-l-,l+1}$, and similarly
for $\hat b',\hat b'^\dag$. 
The $\epsilon$ regularization is introduced for the (finite $N$) singularity created when $\hat{b}^{\dagger}$ ($\hat{b}'^{\dagger}$) acts on $\left|0,N\right>$ ($\left|N,0\right>$), which is singular because there is no particle to be transferred to mode 1 (mode 0) for this state.

We demonstrate in the following that a twofold fragmented state can
be written as the superposition of truncated coherent states of the ladder operators. 
These coherent states live in a finite-dimensional Hilbert space \cite{Kuang,Mira} for 
particle-number-conserved states of the form \eqref{origPsi}.  
The ladder operators approximately satisfy bosonic commutation relations, due to 
$1-\bra{\Psi}[\hat{b},\hat{b}^{\dag}]\ket{\Psi} 
=(N+1)|C_N|^2$ and $1-\bra{\Psi}[\hat{b}',\hat{b}'^{\dag}]\ket{\Psi} =(N+1)|C_0|^2$, 
provided 
$|C_0|,|C_N|\ll 1/\sqrt{N}$.

The truncated coherent states $\ket\beta$ of $\hat b$ 
are defined by  
\begin{multline} 
\label{EVbeta}
\!\hat{b}\ket{\beta} = \hat b \left(A_\beta\sum_{l=0}^N\frac{\beta^l}{\sqrt{l!}}\ket{l}\right)
=\beta\ket{\beta}-\beta A_\beta \frac{\beta^N}{\sqrt{N!}}\ket{N}, 
\end{multline}
where $\ket{l} := \ket{N-l,l}$.
The normalization factor is given by 
$
A_\beta=\exp(-|\beta|^2/2)\sqrt{\frac{\Gamma(N+1)}{\Gamma(N+1,|\beta|^2)}},
$
with the upper incomplete gamma function
$
\Gamma(s,x) = \int_x^{\infty} t^{s-1}\,e^{-t}\,{\rm d}t $.
The state $\ket\beta$ equals the usual  coherent state in a infinite-dimensional Hilbert space 
when for $N\rightarrow \infty$ the second term on the right-hand side of \eqref{EVbeta} 
becomes negligible; 
in fact $\ket{\beta}$ becomes very close to 
a proper coherent state 
already for moderate values of $N$ 
\cite{suppl}.  
Similar considerations hold for the truncated coherent states of the pair 
$\hat b'$, $\ket{\beta'}$. 
Hence both $\ket\beta$ and $\ket{\beta'}$ represent a {finite-$N$ 
generalization} of the concept of coherent states.
The quasiparticles created by $\hat b^\dag$ or, as an 
equivalent choice (also see below), by $\hat b'^\dag$,  
will thus assume the role of the ``photons."

The fragmented condensate 
states are transparently described in terms of phase space
states \cite{Yoo,Castin}, where the relevant phase is conjugate to the occupation number 
difference of the modes. 
For large $N$, fragmented condensates 
correspond to two phase space states  $\ket{\phi,N,l_0}\propto \left(\sqrt{N-l_0}\hat{a}^{\dag}_0+e^{i\phi}\sqrt{l_0}\hat{a}^{\dag}_1\right)^N\ket{{\rm vac}}$, 
separated in phase $\phi$ by exactly $\pi$ \cite{Kang}. 
We now show that individual phase-space basis elements are well approximated by truncated coherent states of 
$\hat b$ $(\hat b')$; we will then conclude that the fragmented ground state can be expressed as a superposition of 
antipodal coherent states, Eq.\,\eqref{mattercat} below.

According to \cite{Kang}, a two-mode many-body state \eqref{origPsi} can 
be expressed in terms of a phase-space basis as
\begin{equation}\label{phasestates}
\ket{\Psi}=\int^{2\pi}_0 \, \frac{d\phi}{2\pi} \, \sum_{l}\mathcal{N}_{N,l_0;l} \,C_l e^{-il\phi}\ket{\phi,N,l_0}, 
\end{equation}
where the relation of the phase space basis states centered at $l_0$  to Fock space basis states is given by 
$\ket{l}=\int^{2\pi}_0 \, \frac{d\phi}{2\pi} \, \mathcal{N}_{N,l_0;l} \, e^{-il\phi}\ket{\phi,N,l_0}$, 
with the normalization factor $\mathcal{N}_{N,l_0;l}=\sqrt{\frac{N^N}{(N-l_0)^{N-l}l_0^l}}\sqrt{\frac{(N-l)!l!}{N!}}$.
Inverting the relation \eqref{phasestates}, 
we have $\ket{\phi,N,l_0}=\sum^N_{l=0} \frac{1}{\mathcal{N}_{N,l_0;l}}e^{il\phi}\ket{l}$, 
and with \eqref{EVbeta}, defining
 $\beta = |\beta| \exp[i\phi_\beta]$, 
\begin{equation}
\ket{\phi,N,l_0}=\frac{1}{A_{\beta}}\int^{2\pi}_0 \frac{d\phi_{\beta}}{2\pi} \sum_l \, \frac{e^{-il(\phi_{\beta}-\phi)}}{\mathcal{N}_{N,l_0;l}} \frac{\sqrt{l!}}{|\beta|^l}\ket{\beta}.
\end{equation}
Now, by setting $|\beta|^2 = l_0$, that is by choosing the mean ``photon" number 
to be equal to the $C_l$-distribution center, 
we obtain 
$ 
\ket{\phi,N,l_0} 
=\sqrt{\frac{N!}{N^N}}\frac{1}{A_{\beta}} \int^{2\pi}_0 \frac{d\phi_{\beta}}{2\pi} \left(\sum_l \sqrt{\frac{(N-l_0)^{N-l}}{(N-l)!}} e^{-il(\phi_{\beta}-\phi)}\right)\ket{\beta}.
$ 
The factor $(N-l_0)^{N-l}/(N-l)!$ is proportional to a Poisson distribution centered around $N-l_0$, well approximated by a normal distribution with both mean and variance $N-l_0$. 
Then 
$ 
\sum^N_{l=0} \sqrt{\frac{(N-l_0)^{N-l}}{(N-l)!}} e^{-il(\phi_{\beta}-\phi)} \propto \sum^N_{l=0} e^{-\frac{(l-l_0)^2}{4(N-l_0)}} e^{-il(\phi_{\beta}-\phi)}.
$ 
This resembles a discrete Fourier transform of the normal distribution 
except that $l\ge 0$ and the summation range is finite. 

We thus find, with the choice $|\beta|^2 = l_0$, 
that a phases-space basis state maps to a truncated coherent  state
of the ladder operators, i.e.,  $\ket{\phi,N,l_0}\simeq\ket{\beta}$, as long as $|\beta|$ and hence 
the fragmentation do not become 
too small, in which case the details of the $C_l$ distribution at the  boundary $l=0$ matter 
and the concept of a $\beta$-coherent state breaks down. 
An analogous argument can be applied to $\ket{\beta'}\simeq \ket{\phi,N,l_0}$
 \cite{suppl}. 
 Below, we restrict ourselves to the truncated coherent states
 associated to the ladder operator $\hat b$, with analogous conclusions holding for $\hat b'$.
 
Using $\ket{\beta}\simeq \ket{\phi,N,l_0}$ and $\ket{-\beta}\simeq \ket{\phi+\pi,N,l_0}$, we thus make the ansatz
$\ket{\Psi}\simeq C_{\beta}\ket{\beta} + C_{-\beta}\ket{-\beta}$
for the fragmented condensate many-body state. 
To identify the coefficients, we observe that 
according to \cite{Bader,Lee}, a fragmented state can equivalently be written as a superposition  of two (for large $N$) 
degenerate many-body states 
\begin{align}
\ket\Psi 
= c \left(\ket{\rm even } + u\exp[i\theta_{\rm K}]  
\ket{\rm odd}\right) , \label{kangsoo} 
\end{align} 
where $\ket{\rm even}$ only contains even $l$ coefficients $C_l$ from the state \eqref{origPsi}   
and $\ket{\rm odd}$ only odd $l$, and 
$|c|^2(1+|u|^2)=1$.
By \eqref{EVbeta} and the fact that $\phi_l ={\rm arg} (C_l)
= \textrm{arg} \left( C_{\beta}  +  C_{-\beta} (-1)^l \right) + \frac{l\pi}{2}$ (for $\phi_{\beta}=\pi/2$), it is established by straightforward algebra that we can write 
$\ket{\rm even}
\simeq\frac{A_{\beta}}{\sqrt{2(1+e^{-2|\beta|^2})}} \sum_{l=0}^N \frac{|\beta|^l}{\sqrt{l!}} \left( i^l (1+(-1)^l) \right) \ket{l}$; further,  $ \ket{\rm odd} \simeq \frac{A_{\beta}}{i\sqrt{2(1-e^{-2|\beta|^2})}} \sum_{l=0}^N \frac{|\beta|^l}{\sqrt{l!}} \left( i^l (1-(-1)^l) \right) \ket{l}$. 
The overlap  
$\braket{-\beta}{\beta}=A_{\beta}^2 \sum_{l=0}^N \frac{(-|\beta|^2)^l}{l!} \simeq e^{-2|\beta|^2}$ for large $N$, 
and we obtain even and odd superpositions of $\ket{\beta}$ and $\ket{-\beta}$, 
$\ket{\rm even} \simeq \frac{1}{\sqrt{2(1+e^{-2|\beta|^2})}} \left( \ket{\beta}+\ket{-\beta} \right)$,  
$\ket{\rm odd} \simeq \frac{1}{i\sqrt{2(1-e^{-2|\beta|^2})}} \left( \ket{\beta}-\ket{-\beta} \right)$ \cite{suppl}. 

In summary, we have established that a fragmented condensate 
state 
can be written as a superposition of ``photonic" truncated coherent states of the ladder operators defined in \eqref{def b}, 
\bea
\ket{\Psi} = \mathcal{N} (\ket{\beta}+r e^{i\theta}\ket{-\beta}), \qquad |\beta|^2=l_0, \label{mattercat} 
\ea 
where 
$\mathcal N 
=\left( 1 + r^2+ 2r\cos\theta \exp[-2|\beta|^2]
\right)^{-1/2}$. 
The phase $\theta$ and coefficient $r$ in terms of $u,\theta_{\rm K}$ in \eqref{kangsoo}  
are found from
$r\exp[i\theta]=\frac{1+ u \lambda_{\beta} e^{i(\theta_k+\pi/2)}}{1-u \lambda_{\beta} e^{i(\theta_k+\pi/2)}} $, where $\lambda_{\beta}=\sqrt{\frac{1+e^{-2|\beta|^2}}{1-e^{-2|\beta|^2}}}$.
In the limit $|\beta|^2\gg 1$, 
\bea
\!\!\!\!\! r = \left| \frac{ (1-u^2) + 2iu\cos\theta_{\rm K} }{ (1+u^2) + 2u\sin\theta_{\rm K} } \right|\!,\,\,     
\theta = 
\tan^{-1} \left( \frac{2u\cos\theta_{\rm K}}{1-u^2} \right)\!. 
\ea
We thus find, in the equal-weight case $r=1$, that  
the fragmented two-mode many-body states  represent the
analogue of photon cat states \cite{Ourjoumtsev,Afek,Gao,Yao}.
For the latter, the truncated coherent state 
$\ket\beta$ of the massive bosons is replaced by 
the coherent state of, e.g., a cavity photon field. 

It was shown in \cite{Xiong} that a stable variety of  fragmented two-mode many-body states is obtained for $r=1$, $\theta=\pi/2,3\pi/2$.
In the quantum optics 
literature on photon cat states, due to their potential for creating entangled coherent states, 
cf. the review \cite{Sanders},  the 
states $\ket{\rm even},\ket{\rm odd}$ are
frequently studied,   
which correspond to $r=1$, $\theta =0 ,\pi$. 
They are susceptible to perturbations when $|\beta|^2\gg 1$, which has been borne out for the presently studied "photonic" case in an interacting
gas of massive bosons \cite{Xiong}. 
Finally, $r\ll 1$ and $r\gg 1$ yield 
$\ket{\beta}$ and $\ket{-\beta}$, respectively, which represent 
nonfragmented condensation. 

The variances of effective 
position and momentum variables (quadratures), in ladder space, are obtained  from 
\bea
\!\pm \Delta (\hat b \pm \hat b^\dag)
=  \begin{cases} 4|\beta|^2 \cos^2\phi_{\beta}\left( \frac{2r}{1+r^2} \right)^2 
+1
,\; ``+"\,,\\
4|\beta|^2 \sin^2\phi_{\beta}\left( \frac{2r}{1+r^2} \right)^2 +1 
,\; ``-"\,  ,
\end{cases}
\label{quad}
\ea
for $|\beta|^2\gg 1$, and where $\Delta (\hat b \pm \hat b^\dag);=  
\langle(\hat{b}\pm\hat{b}^{\dag})^2 \rangle 
- \langle\hat{b}\pm\hat{b}^{\dag}\rangle^2$.  
The superposition \eqref{mattercat} 
is a proper cat state for $r\rightarrow 1$, 
with large quantum fluctuations of  ``position" and ``momentum." In \eqref{quad}, 
$\phi_\beta$ can be considered as a parameter determining in which direction 
of ``photonic" quadrature fluctuations are squeezed.

The size of the SIMCAS (that is their degree of quantum mechanical superposition macroscopicity) 
is, in accordance with the quadratures \eqref{quad}, determined by 
 $|\beta|^2$. 
 The size of coherent state superpositions such as \eqref{mattercat}
 is maximal when the 
 overlap $\braket{-\beta}{\beta}\simeq \exp[-2|\beta|^2]$ 
  is minimal \cite{Duer,Volkoff}, the size $\mathcal M$ 
 being a simple polynomial function of the overlap, ${\mathcal M} \simeq (1-\exp[-2|\beta|^2])^2$. 
 Thus, due to $|\beta|^2={\mathcal F}N/2$,  
 \bea
{\mathcal M} \simeq 
(1-\exp[-{\mathcal F}N])^2\qquad (r=1). \label{catsize}
 \ea 
 The exponential dependence of the superposition size $\mathcal M$ on $\mathcal F$ highlights the distinctly nonclassical 
 character of a fragmented condensate in comparison to a BEC (single condensate). 
 This will be manifest already for relatively small values of $\mathcal F$, when $N$ is mesoscopic, as
  in experimentally realized quantum gases.
 We remark 
 that there exist alternative measures for 
 assessing the macroscopicity of superpositions, which rely on calculating the quantum Fisher information \cite{DuerNJP,Volkoff2}, but obviously also involve the overlap 
 $\braket{-\beta}{\beta}$ in an essential manner.  

 In marked distinction to superpositions of photon coherent states, 
 the superposition size $\mathcal M$ 
strongly depends not only on a particle number,  
$N$, but also, via $\mathcal F$, on the strength of the microscopic (two-body) interactions 
in the constituent system dilute quantum gas; $\mathcal M$ in fact vanishes exponentially fast for weakly interacting systems where $\mathcal F\rightarrow 0$. 
The macroscopic superposition size  
of the SIMCAS is therefore a genuine many-body effect \cite{note}. 
We note in this regard that the macroscopicity $\cal M$ for our matter wave cat is assessed by the rather straightforward 
measurement of density-density correlations (see below), and not the elaborate 
quantum state tomography required for proper photonic cat states.

Next we establish the relation of the ``photonic" quadratures \eqref{quad} 
to the density-density correlations in the interacting quantum gas. 
The nonclassical character of the superposition 
\eqref{mattercat}, in conjunction with its many-body origin, will, by this means,  
become experimentally verifiable. 
We consider the 
density-density correlations in an effectively one-dimensional (1D) system (also see the below discussion
on the experimental implementation),  
$\Delta\rho_2(z,z'):= \big< \hat \rho(z) \hat \rho (z') \big>
-\rho_1(z)\rho_1(z')$, where the density 
$\rho_1(z)=\big<\hat\psi^\dag(z)\hat\psi(z)\big>$. We take the approximation that 
$ \big< \hat \rho(z) \hat \rho (z') \big>
\simeq \big<\hat{\psi}^{\dagger}(z)\hat{\psi}^{\dagger}(z')\hat{\psi}(z')\hat{\psi}(z)\big>$, 
which holds true for $\big< \hat N_i \big> \gg 1$. 
For simplicity, we assume that the two macroscopically occupied 
field operator modes have even and odd parity, respectively, $\hat \psi(z) =
\psi_0(z) \hat a_0 + \psi_1(z) \hat a_1$, with $\psi_0(z)=\psi_0 (-z)$ and $\psi_1(z) = -\psi_1(-z)$, 
and 
are real, $\psi_i(z)\in \mathbb{R}$.
Bearing in mind that a fully self-consistent solution of the many-body Schr\"odinger equation will
yield the true orbitals cf., e.g. \cite{Alon,Alon2,Chatterjee,IJMPB}, for illustration purposes we take them 
to be the ground and first excited states of the harmonic oscillator; $\psi_0(z)=\pi^{-1/4}\exp\left[-{z^2}/{2}\right]$ and $\psi_1(z)=\pi^{-1/4} \sqrt 2 z \exp\left[-{z^2}/{2}\right]$, where the $z$ coordinate 
is assumed to be scaled by a suitable measure of (half the) extension of the cloud.

For $|\beta|^2\gg 1, r=1$, 
we obtain 
the correlation function  
$\Delta\rho_2(z,z')= \psi_0(z)\psi_1(z')\psi_0(z')\psi_1(z)
[ 2 \big< \hat{a}_0^{\dag} \hat{a}_1^{\dag} \hat{a}_1 \hat{a}_0 \big> + \big< \hat{a}_0^{\dag} \hat{a}_0^{\dag} \hat{a}_1 \hat{a}_1 \big> + \big< \hat{a}_1^{\dag} \hat{a}_1^{\dag} \hat{a}_0 \hat{a}_0 \big>
- ( \big< \hat{a}_0^{\dag} \hat{a}_1 + \hat{a}_1^{\dag} \hat{a}_0 \big> )^2 ]$. 
Converting this 
to ``photonic" ladder space, we get 
\begin{multline}
\hspace*{-1em}
\Delta \rho_2 (z,z') \simeq \psi_0(z)\psi_1(z')\psi_0(z')\psi_1(z)\\
%
\times\left[ 
\left\langle \left( {\hat{N}_0}^{1/2} \, (\hat{b}+\hat{b}^{\dag}) \right)^2 \right\rangle 
- \left( 
\left\langle 
{\hat{N}_0}^{1/2} 
\, (\hat{b}+\hat{b}^{\dag}) 
\right\rangle 
\right)^2 
\right] .
\label{dd correl}
\end{multline}
The 
density-density correlations in the strongly interacting gas therefore map out the quadrature in the  first line of \eqref{quad} when the fluctuation of 
occupation numbers is small, 
i.e. for large $N$ at finite $\mathcal F$.

\begin{center}
\begin{figure}[t]
\hspace*{-0.8em}
\includegraphics[width=1.02\columnwidth]{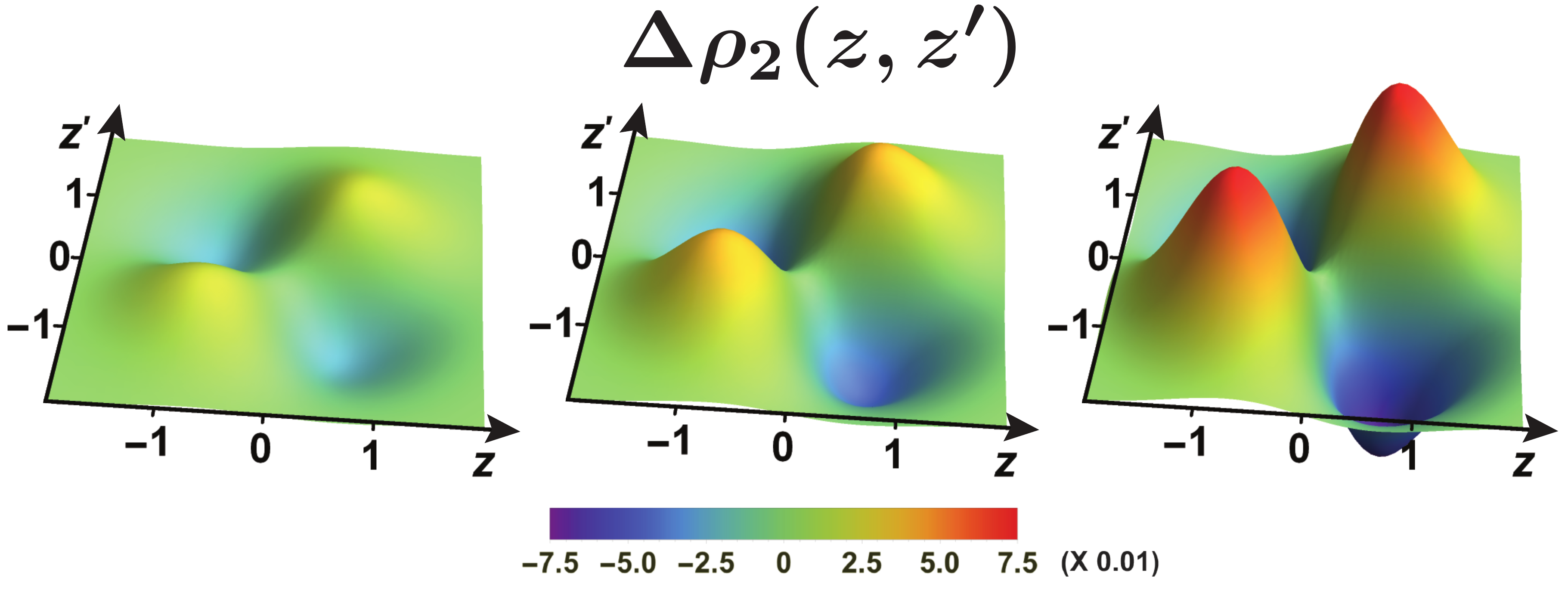}\vspace*{1em}
\includegraphics[width=0.9\columnwidth]{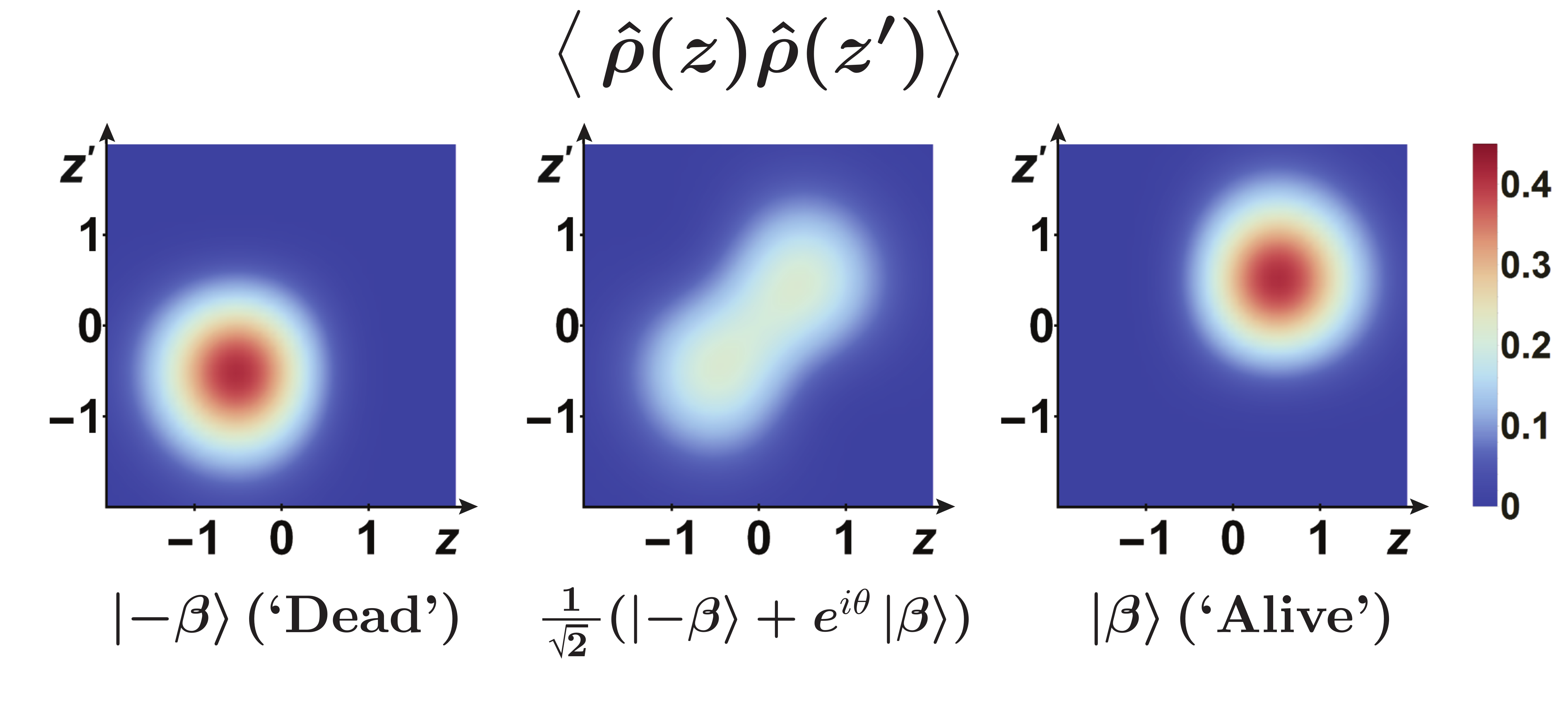}
\caption{\label{fig1} Emergence of a ``photonic" 
cat state superposition 
in an ultracold quantum gas containing 
$N=100$ massive, interacting bosons. 
Top: 1D density-density correlations $\Delta \rho_2(z,z')$,   
scaled by $N^2/Z^2$, 
where $Z$ is 
the extension of the cloud after axial TOF ($Z\propto t$ in the long time limit \cite{Kang}), 
which is also the scaling unit of $z,z'$. 
From left to right, the fragmentation degree increases according to 
${\mathcal F}=0.1,0.2,0.4$ ($|\beta|^2 
=5,10,20$, with $r=1$). 
Bottom: Density-density correlator 
$\big<\hat \rho(z)\hat \rho(z')\big>$ (in the same $N^2/Z^2$ units) 
for $|\beta|^2=20$, and 
$r\rightarrow 0$ (left), $r=1$ (middle), $r\rightarrow \infty$ (right). The correlator does not depend on $\theta$ in the large $|\beta|^2$ limit.} 
\end{figure}
\end{center}

Before TOF, $\phi_\beta=\frac\pi2$, 
with no distinct correlation signal present (the negative pair coherence of a fragmented state  
enforces sgn$(C_{l+2}C_l)=-1$  \cite{Bader,Lee,Kang}, which in turn implies the condition 
$2\phi_{\beta}=\pi+2n\pi$, where $n$ is an integer \cite{suppl}).
After expansion of the cloud, it has been shown in \cite{Kang} that TOF of the cloud in the axial direction effectively performs a rotation $\phi_\beta 
\rightarrow \phi_\beta-\pi/2$ of the relative phase of the two modes, 
cf. the definition of $\hat b$ in (2) involving $\hat a_0^\dag\hat a_1$. This results in $\phi_\beta=0$, and we conclude from (9) and (11) that a characteristic correlation signal develops \cite{squeezing}.
We plot the density-density correlations according to the state \eqref{mattercat}
in Fig.\,\ref{fig1}, which maps out with increasing accuracy the exact 
correlations of the underlying fragmented many-body state \eqref{kangsoo} when $|\beta|^2$ increases \cite{suppl}.
One clearly recognizes the emerging 
strong correlation signature  of an increasingly macroscopic 
SIMCAS for larger ``photon" number $|\beta|^2$: The larger the 
degree of fragmentation, the bigger the cat becomes.
To identify the `dead' and `alive' parts of the superposition state, we plot 
in the bottom part of Fig.\,\ref{fig1} 
the density-density correlator
$\big<\hat \rho(z)\hat \rho(z')\big>$,  
for three limiting cases of superposition weights in \eqref{mattercat}, 
$r=0,1,\infty$, yielding $\ket{\beta}$, $\frac1{\sqrt2}(\ket{\beta}
+\exp[i\theta] \ket{-\beta})$, and $\ket{-\beta}$, respectively. 

For the experimental realization of SIMCAS, quasi-1D  systems \cite{LiebLiniger} 
are favorable, since they possess larger degrees of fragmentation when compared to higher-dimensional gases \cite{Chatterjee}. 
SIMCAS  exist in a interaction coupling--density region of the phase diagram 
relatively close to the BEC domain and far from the extreme Tonks-Girardeau 
regime of ultralow density and very large coupling
 \cite{Dunjko,Paredes} (in which the number of fragments equals the number of particles \cite{AlonPRL}). 
 %
 The requirements on gas density and contact interaction coupling constant $g$ to realize SIMCAS
are therefore more moderate than for the Tonks-Girardeau gas, and should be readily accessible in experiment.
Finally, the phase $\theta_{\rm K}$ and amplitude $r$ in the superposition \eqref{kangsoo}
can be engineered by 
a rapid sweep of interaction couplings, as 
demonstrated 
for $\theta_{\rm K}$ in  
\cite{Lee}. 

\vspace*{-3em} 

Our analysis reveals the highly nonclassical 
character of fragmented condensate many-body states, 
as opposed to the essentially classical BEC contained in the same trap.  
We therefore anticipate applications of SIMCAS, 
{\it inter alia}, in quantum metrology  \cite{Giovannetti}, where
{the Cram\'er-Rao bound  
furnishes rigorous bounds on the precision to which parameters in the  %
Hamiltonian 
can be measured \cite{ParisReview}. 
The quantum superposition macroscopicity 
of SIMCAS manifest in the density-density correlations of the bosonic gas 
thus establishes a potential
many-body resource of parameter estimation theory. 


We thank Ty Volkoff and Daniel Braun for helpful discussions. 
This research was supported by the NRF Korea, Grant No. 2014R1A2A2A01006535.



\begin{widetext}
\setcounter{equation}{0}
\setcounter{figure}{0}
\setcounter{table}{0}
\renewcommand{\theequation}{S\arabic{equation}}
\renewcommand{\thefigure}{S\arabic{figure}}

\section{Supplemental Material}
We here provide a more detailed analysis of both the accuracy of 
using $\ket{\beta}$ as approximate (truncated) coherent states, as well as the accuracy of the identification of phase-space basis states with them,  
$\ket{\phi,N,l_0}\simeq\ket{\beta}$. Furthermore, 
a more extended description of the TOF evolution of density-density correlations is presented, and the 
accuracy of the cat state ansatz $\ket{\Psi}=\mathcal{N}(\ket{\beta}+r^{i\theta}\ket{-\beta})$
for the fragmented condensate many-body state is assessed.

\subsection{ Accuracy of Treating $\ket{\beta}$ as a Coherent State}

In the main text, we introduced in Eq.\,(2) ladder operators, which represent approximate 
bosonic annihilation operators, as follows 
\begin{equation}
\hat{b}= \lim_{\epsilon \rightarrow 0} \frac{1}{\sqrt{\hat{N}_0+\epsilon}}\hat{a}^{\dag}_0\hat{a}_1, \quad
\hat{b}'=\lim_{\epsilon \rightarrow 0} \frac{1}{\sqrt{\hat{N}_1+\epsilon}}\hat{a}^{\dag}_1\hat{a}_0, 
\end{equation}
and the corresponding truncated coherent states as 
\begin{equation}
\ket{\beta}=A_\beta\sum_{l=0}^N\frac{\beta^l}{\sqrt{l!}}\ket{N-l,l}, \quad \ket{\beta'}=A_{\beta'}\sum_{l=0}^N\frac{\beta'^{N-l}}{\sqrt{(N-l)!}}\ket{N-l,l},
\end{equation}
where $A_{\beta}=\exp(-|\beta|^2/2)\sqrt{\frac{\Gamma(N+1)}{\Gamma(N+1,|\beta|^2)}}$.
Truncated coherent states were previously treated, e.g., by \cite{Kuang}. 
They are finite-dimensional-Hilbert-space versions of the bosonic annihilation operators 
$\hat{a}$ and coherent state $\ket{\alpha}$ in quantum optics. 
Here, we aim at finding a quantity which assesses the accuracy of treating $\hat{b}$ and $\ket{\beta}$ ($\hat{b'}$ and $\ket{\beta'}$) as bosonic annihilation operators and their corresponding eigenstates,
\begin{equation} \label{b-condition}
[\hat{b},\hat{b}^{\dag}]=1,\quad \hat{b}\ket{\beta}=\beta\ket{\beta}, 
\end{equation} 
and which allows us to numerically evaluate the precision to which the above two relations hold. 

We assume, without loss of generality, $\big<\hat{a}_1^{\dag}\hat{a}_1\big>\leq N/2$ with $|\beta|^2\leq N/2$ because, as will be shown in the following, more robust sets of bosonic operator and their eigenstate are $\{\hat{b},\ket{\beta}\} $ when $\big<\hat{a}_1^{\dag}\hat{a}_1\big>\leq N/2$, and $\{\hat{b'},\ket{\beta'}\}$ when $\big<\hat{a}_1^{\dag}\hat{a}_1\big>\geq N/2$. In addition, we note that 
there is always the freedom to choose $\hat{b},\ket{\beta}$ or $\hat{b'},\ket{\beta'}$ to describe a given two mode system.
With respect to an arbitrary (normalized) two-mode state $\ket{\Psi}$
\begin{equation}
\ket{\Psi}=\sum_{l=0}^N C_l\ket{N-l,l}=\sum_{l=0}^N C_l \frac{(\hat{a}_0^{\dag})^{N-l}(\hat{a}_1^{\dag})^l}{\sqrt{(N-l)!\,l!}}\ket{0},~ \sum_{l=0}^N |C_l|^2=1, 
\end{equation}
the accuracy of \eqref{b-condition} can be assessed by evaluating the following two quantities
\begin{equation} \label{cond1}
1-\bra{\Psi}[\hat{b},\hat{b}^{\dag}]\ket{\Psi}=(N+1)|C_N|^2 , \quad
\frac{\left(\bra{\beta}\beta^*-\bra{\beta}\hat{b}^{\dag}\right) \left(\beta\ket{\beta}-\hat{b}\ket{\beta}\right)}{ |\beta|^2\braket{\beta\,}{\beta}} = A_{\beta}^2 \frac{|\beta|^N}{N!}.
\end{equation}
It is readily observed that the accuracy of $\hat{b},\hat{b}^{\dag}$ as a proper set of bosonic operators depends on the state $\ket{\Psi}$ which is considered. In the main text a twofold fragmented state is discussed with the ansatz $\ket{\Psi}= \mathcal{N}(\ket{\beta}+r^{i\theta}\ket{-\beta})$ (also see below), and since the $|C_N|^2$ distributions of $\ket{\beta}$ and $\ket{-\beta}$ are identical, it is sufficient 
to consider $\ket{\Psi}=\ket{\beta}$.
Eq.\,\eqref{cond1} becomes
\begin{equation}
1-\bra{\beta}(\hat{b}\hat{b}^{\dag}-\hat{b}^{\dag}\hat{b})\ket{\beta}=(N+1) A_{\beta}^2 \frac{|\beta|^{2N}}{N!} , \quad
\frac{\left(\bra{\beta}\beta^*-\bra{\beta}\hat{b}^{\dag}\right) \left(\beta\ket{\beta}-\hat{b}\ket{\beta}\right)}{|\beta|^2 \braket{\beta\,}{\beta}} = A_{\beta}^2 \frac{|\beta|^{2N}}{N!}.
\end{equation}
Higher order expressions in $\hat{b},\hat{b}^{\dag}$ 
such as $\hat{b}\hat{b}\hat{b}^{\dag}\hat{b}^{\dag}$ or the repetitive action of $\hat{b}$ against $\ket{\beta}$, can be similarly assessed.
They induce the ``loss" of certain amplitudes $C_{N-1},\cdots,C_{N-n+1}$ from the state vector 
$\ket{\beta}$, where
\begin{equation} \label{apcoh_n}
\beta^n\ket{\beta}-(\hat{b})^n\ket{\beta}=\beta^n A_{\beta} \sum_{l=N-n+1}^N \frac{\beta^l}{\sqrt{l!}}\ket{N-l,l}.
\end{equation} 
 Therefore the {\em robustness} of the coherent state representation 
can be defined by the degree to which a state of interest $\ket{\Psi}$ is not deformed 
from the loss of the amplitudes $C_N,\cdots,C_{N-n+1}$ due to a $n$-times repeated action 
of $\hat{b}$ on it.
Specifically, for $\ket{\Psi}=\ket{\beta}$ this robustness can be quantified through  
\begin{equation} \label{cond2}
\frac{\sum_{l=N-n+1}^{N}\frac{|\beta|^{2l}}{l!}}{\sum_{l=0}^{N}\frac{|\beta|^{2l}}{l!}}=\frac{\Gamma(N+1,|\beta|^2)-\Gamma(N+1)\frac{\Gamma(N-n+1,|\beta|^2)}{\Gamma(N-n+1)}}{\Gamma(N+1,|\beta|^2)} \, \simeq \, \exp(-|\beta|^2)\sum_{l=N-n+1}^{N}\frac{|\beta|^{2l}}{l!},
\end{equation}
which represents the lost fraction of the $C_l$ due to repetitive action of $\hat{b}$. We used here   
$\Gamma(N+1,|\beta|^2)\simeq \Gamma(N+1)$ which holds within $1\%$ error for $N>10$ 
with given $|\beta|^2\leq N/2$. 
In Fig.\,\ref{SMfig1}, we show plots of the quantity defined in \eqref{cond2} for 
different $N,n,|\beta|^2/N$. We see that as $N$ gets larger and $|\beta|^2/N$ gets smaller, the robustness significantly increases. Specifically, for $N=25,|\beta|^2=0.5N$ there is already 
a negligible loss of $C_l$ amplitudes from the state even upon $n=5$ times acting with $\hat{b}$ on it.

\begin{figure} [t]
\begin{center}
\includegraphics[width=.4\columnwidth]{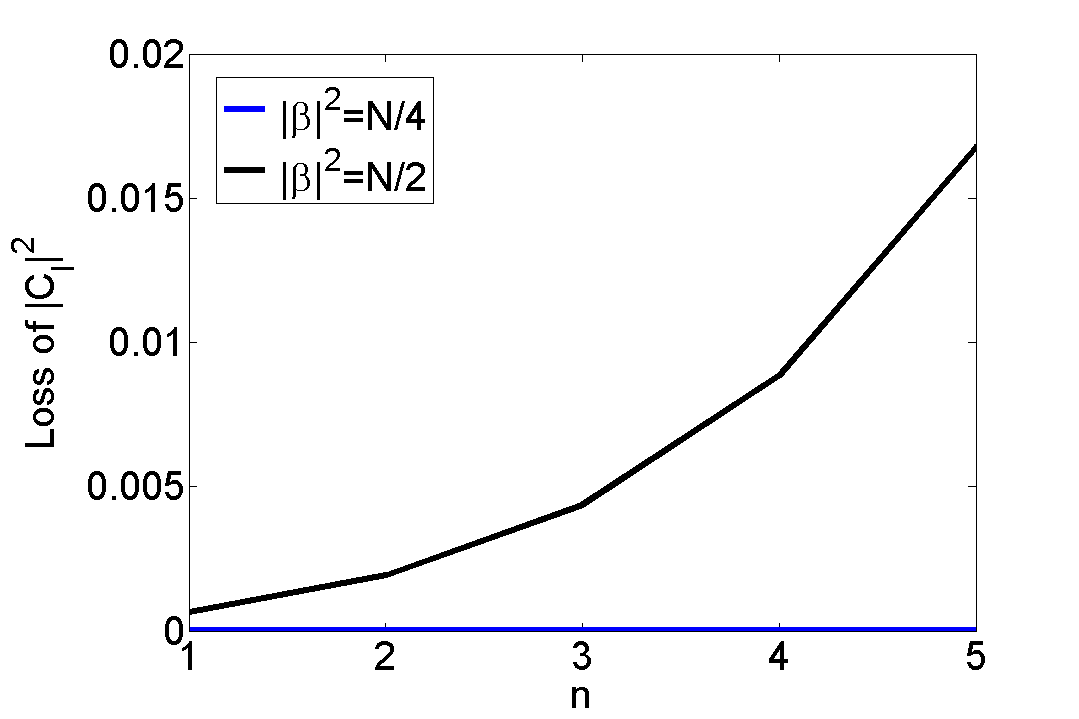} \quad
\includegraphics[width=.4\columnwidth]{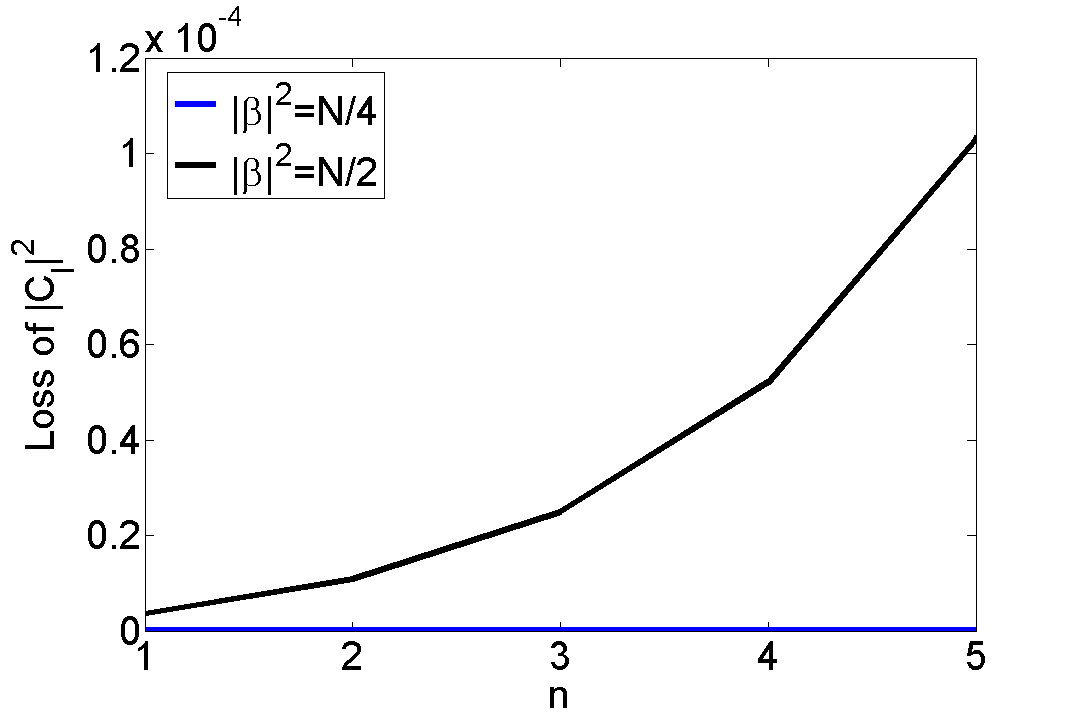}
\end{center}
\caption{\label{SMfig1} Plots of $\frac{\sum_{l=N-n+1}^{N}\frac{|\beta|^{2l}}{l!}}{\sum_{l=0}^{N}\frac{|\beta|^{2l}}{l!}}$  (indicated by loss of $|C_l|^2$ on the vertical axis),  
with $|\beta|^2=N/4$ (blue) and $|\beta|^2=N/2$ (black), for $N=25$ (left) and $N=50$ (right).} 
\end{figure}

\subsection{Comparison of $\ket{\phi,N,l_0},\ket{\beta},\ket{\beta'}$}

\begin{figure} [h]
\begin{center}
\includegraphics[width=.32\columnwidth]{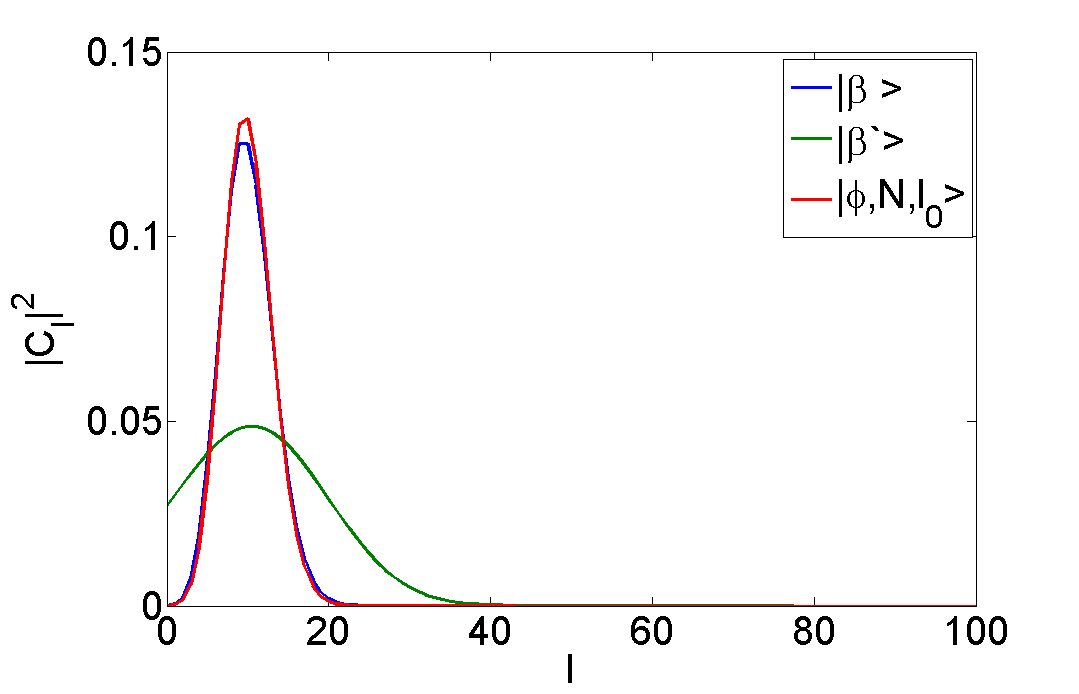}
\includegraphics[width=.32\columnwidth]{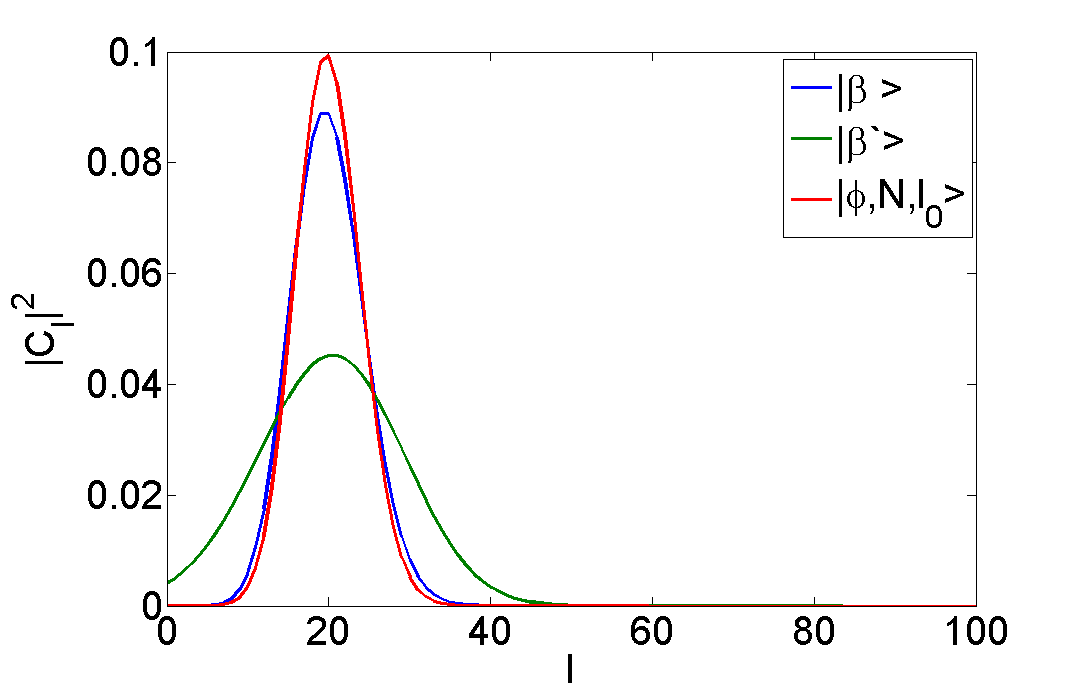}
\includegraphics[width=.32\columnwidth]{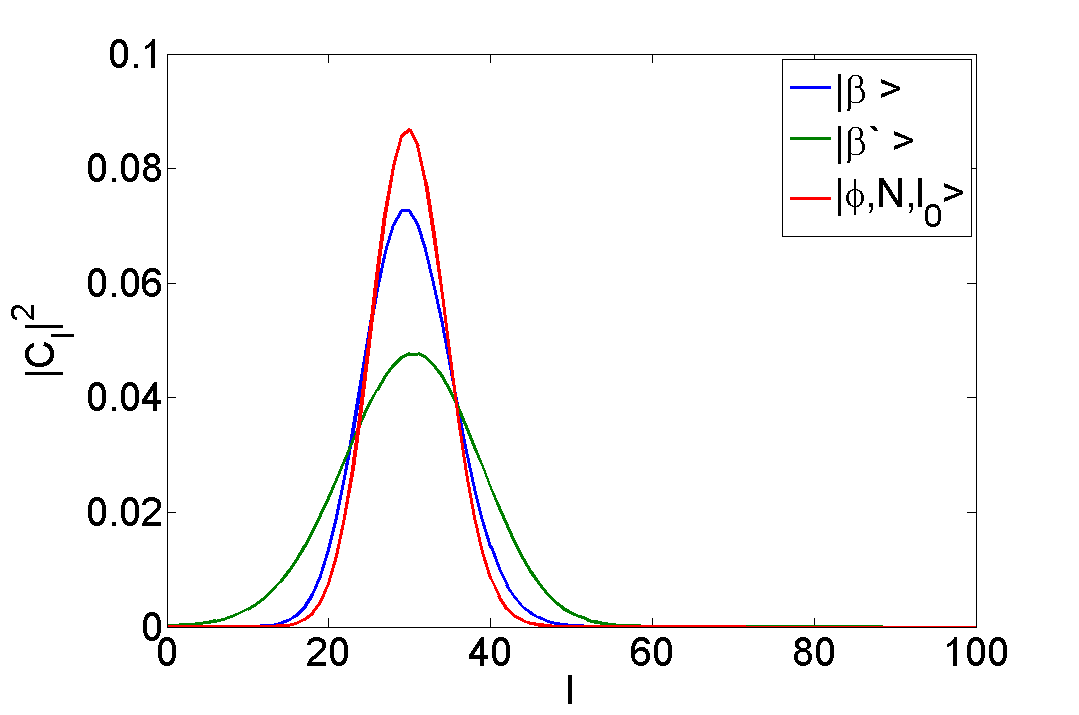} \\
\includegraphics[width=.32\columnwidth]{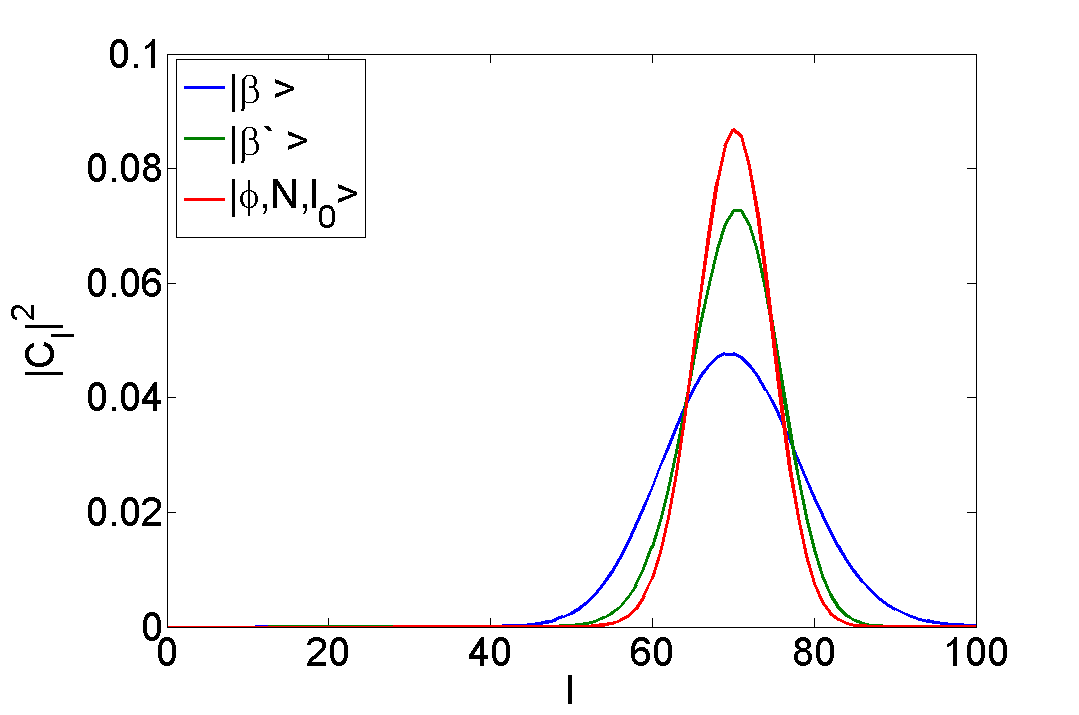}
\includegraphics[width=.32\columnwidth]{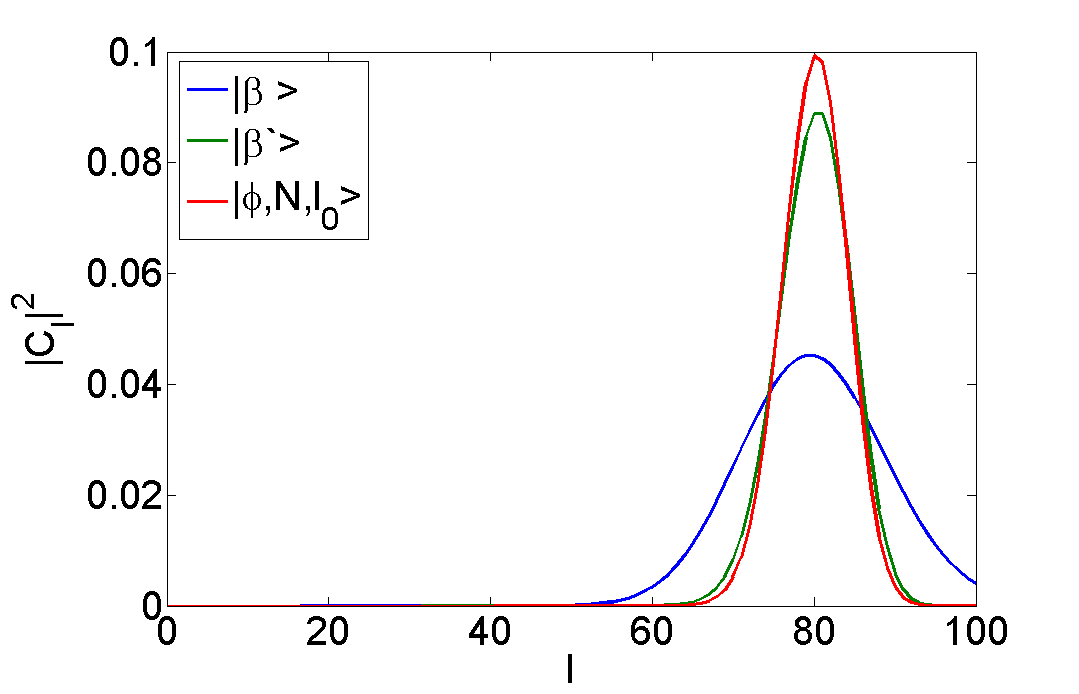}
\includegraphics[width=.32\columnwidth]{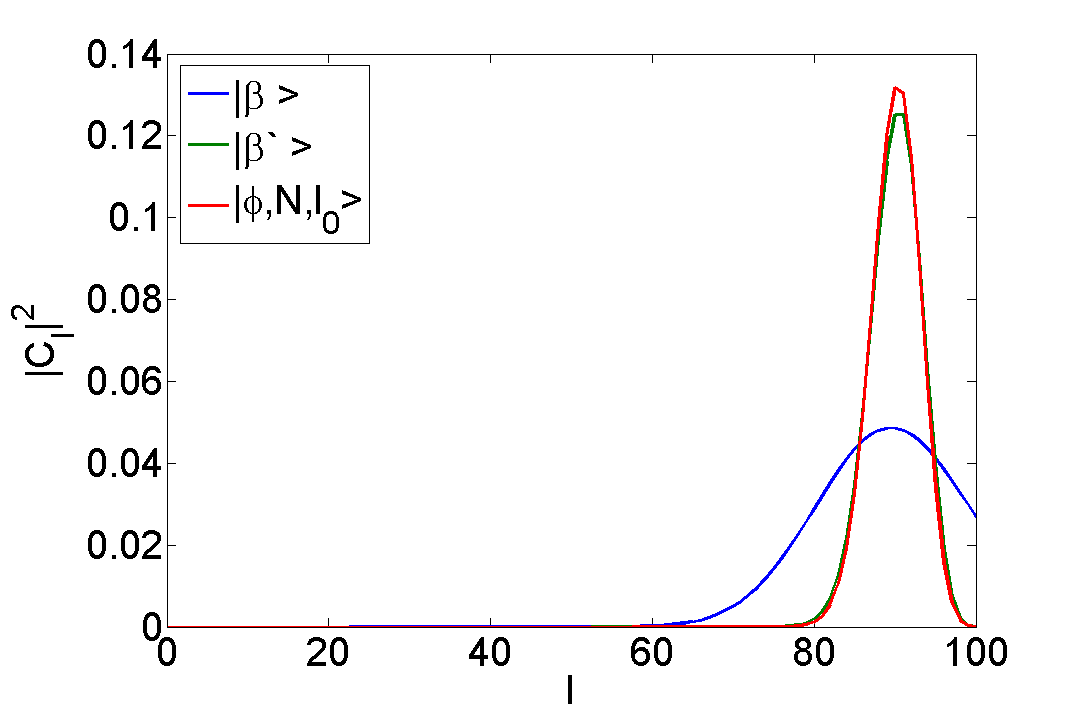}
\end{center}
\caption{\label{SMfig2} Plots of the $|C_l|^2$ distribution of $\ket{\phi,N,l_0}$ (red), $\ket{\beta}$ (blue), $\ket{\beta'}$ (green) with $N=100$ for different $l_0=|\beta|^2=N-|\beta'|^2=0.1N,0.2N,0.3N$ (Top from left) and $l_0=|\beta|^2=N-|\beta'|^2=0.7N,0.8N,0.9N$ (Bottom from left).}
\end{figure}

We argue in the main text that $\ket{\phi,N,l_0}\simeq\ket{\beta}$ for small $l_0=|\beta|^2$, and $\ket{\phi,N,l_0}\simeq\ket{\beta'}$ for small $N-l_0=|\beta'|^2$ with  $\phi=\phi_{\beta}$ and $\phi=-\phi_{\beta'}$, cf.\,the discussion after Eq.\,(5) in the main text and Eq.\,\eqref{relation} below. 
We present in Fig.\,\ref{SMfig2} the $|C_l|^2$ distribution for $\ket{\phi,N,l_0},\ket{\beta},\ket{\beta'}$, respectively, with different $l_0=|\beta|^2$ in the case $N=100$.
We conclude 
that $\ket{\phi,N,l_0}\simeq\ket{\beta}$ is confirmed for small $l_0=|\beta|^2$, and $\ket{\phi,N,l_0}\simeq\ket{\beta'}$ for small $N-l_0=|\beta'|^2$ with  $\phi=\phi_{\beta}$ and $\phi=-\phi_{\beta'}$. 

Furthermore, we have verified that in the large $N$ limit, the truncated coherent states to increasingly good accuracy represent the corresponding phase-space basis vectors even when $l_0 \rightarrow \frac N2$, that is in the limit of maximal fragmentation, ${\mathcal F}\rightarrow 1$. 
In more detail, as $|\beta|^2$ approaches $|\beta|^2=N/2$ ($\mathcal{F}=1$), the $|C_l|^2$ distribution of the phase state $\ket{\phi,N,l_0}$ becomes wider than that of $\ket{\beta}$ or $\ket{\beta'}$. However, $\ket{\phi,N,l_0}\simeq \ket{\beta}$ still holds. From equation (5) of the main text, which transforms $\ket{\phi,N,l_0}$ into the $\ket{\beta}$ basis, we have
\begin{equation}
\ket{\phi,N,l_0} =\sqrt{\frac{N!}{N^N}}\frac{1}{A_{\beta}} \int^{2\pi}_0 \frac{d\phi_{\beta}}{2\pi}
 \left(\sum_l \sqrt{\frac{(N-l_0)^{N-l}}{(N-l)!}} e^{-il(\phi_{\beta}-\phi)}\right)\ket{\beta}=  \int^{2\pi}_0 \frac{d\phi_{\beta}}{2\pi} C_{\phi_{\beta}}  \ket{\beta}. \label{relation}
\end{equation}
with $|\beta|^2=l_0$ where $C_{\phi_{\beta}}=\sqrt{\frac{N!}{N^N}}\frac{1}{A_{\beta}} \left(\sum_l \sqrt{\frac{(N-l_0)^{N-l}}{(N-l)!}} e^{-il(\phi_{\beta}-\phi)}\right)$. 
In Fig.\,\ref{SMfig3}, the modulus of the coherent state phase distribution, $|C_{\phi_{\beta}}|$, is plotted for $l_0=|\beta|^2=N/2$, and with 
various values of particle number $N$.
\begin{figure} [t]
\begin{center}
\includegraphics[width=.5\columnwidth]{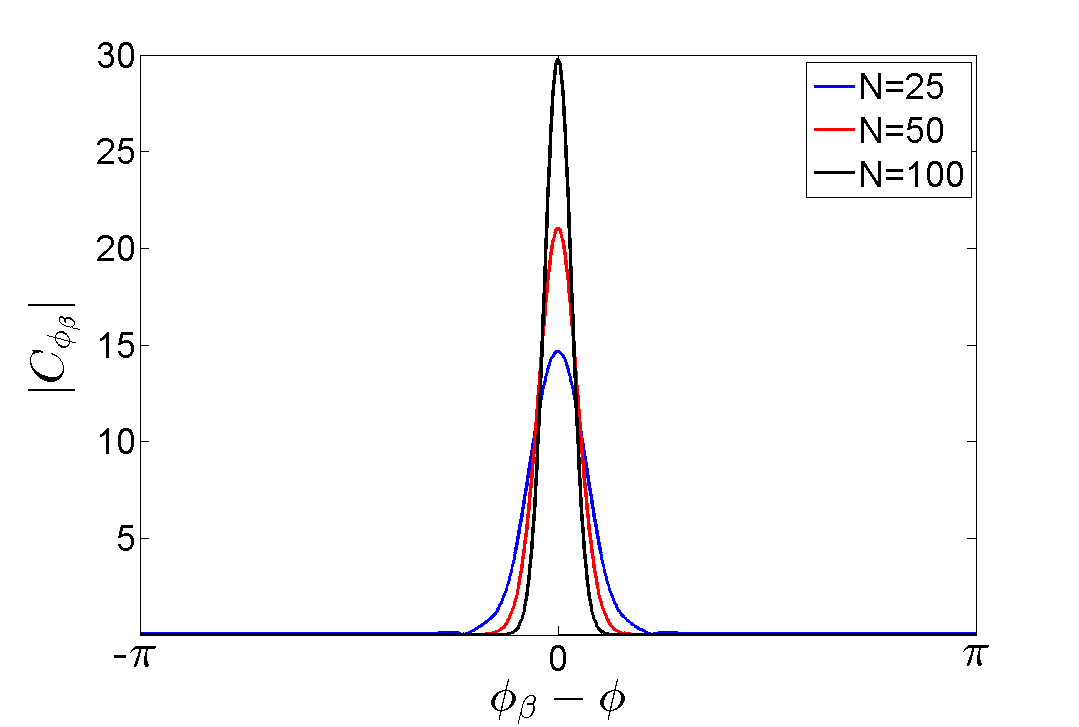}
\end{center}
\caption{\label{SMfig3} Plots of the $|C_{\phi_{\beta}}|$ distribution of $\ket{\phi,N,l_0}$ 
(cf.\,Eq.\,(5) in the main text and the discussion that follows it), for $l_0=|\beta|^2=N/2$ with $N=25,50,100$. 
}
\end{figure}
It is apparent that, with increasing $N$, even when $N=25$, a rapid convergence of $|C_{\phi_{\beta}}|$ to one peak at $\phi_{\beta}=\phi$ occurs. 

\subsection{Time-Of-Flight Expansion (TOF) and 
Phase Rotation}

We present here a more detailed analysis of the phase rotation incurred by TOF.

After TOF, the assumed even-odd parity of the modes is preserved, 
$\psi_0(z,t)=\psi_0(-z,t),\psi_1(z,t)=-\psi_1(-z,t)$. Defining the phase variable 
$\varphi$ such that
\begin{equation}
e^{-i\varphi} \psi_1(z,t) := \bar{\psi}_1(z,t), \quad \psi_0^*(z,t) \bar{\psi}_1(z,t) \in \mathbb{R}
\end{equation}
the wave functions $\psi_0(z,t)$ and $\psi_1(z,t)$ share a 
common phase factor \cite{Gomes}, keeping $\big<\hat{a}_i^{\dag}\hat{a}_j\big>,\big<\hat{a}_i^{\dag}\hat{a}_j^{\dag}\hat{a}_k\hat{a}_l\big>$ ($i,j,k,l=0,1$) invariant under time evolution in the 
noninteracting limit of TOF. Before TOF ($t=0$) $\varphi=0$ and after TOF (large $t$) $\varphi=-\pi/2$ \cite{Kang}.

Note that the value of $\varphi$ could be both $\varphi$ or $\varphi+\pi$ since $e^{i\pi}$ times a real number is again real number. 
We write down the density-density correlation function $\Delta\rho_2(z,z',t)$ as
\begin{equation} \label{densityfluctuationvarphi}
\begin{split}
& \Delta\rho_2(z,z',t) = |\psi_0(z,t)|^2|\psi_0(z',t)|^2 \left(\big< \hat{a}_0^{\dag} \hat{a}_0^{\dag} \hat{a}_0 \hat{a}_0 \big> - \big< \hat{a}_0^{\dag} \hat{a}_0 \big> \big< \hat{a}_0^{\dag} \hat{a}_0 \big> \right)
 + |\psi_1(z,t)|^2|\psi_1(z',t)|^2 \left(\big< \hat{a}_1^{\dag} \hat{a}_1^{\dag} \hat{a}_1 \hat{a}_1 \big> - \big< \hat{a}_1^{\dag} \hat{a}_1 \big> \big< \hat{a}_1^{\dag} \hat{a}_1 \big> \right) \\
&+ \left( |\psi_0(z,t)|^2|\psi_1(z',t)|^2 + |\psi_1(z,t)|^2|\psi_0(z',t)|^2 \right) \left(\big< \hat{a}_0^{\dag} \hat{a}_1^{\dag} \hat{a}_1 \hat{a}_0 \big> - \big< \hat{a}_0^{\dag} \hat{a}_0 \big> \big< \hat{a}_1^{\dag} \hat{a}_1 \big> \right) \\
&+ \left( |\psi_0(z,t)|^2 \psi_0^*(z',t) \bar{\psi}_1(z',t)  + |\psi_0(z',t)|^2 \psi_0^*(z,t) \bar{\psi}_1(z,t)  \right)
\left[ e^{i\varphi} \left(\big< \hat{a}_0^{\dag} \hat{a}_0^{\dag} \hat{a}_0 \hat{a}_1 \big> - \big< \hat{a}_0^{\dag} \hat{a}_0 \big> \big< \hat{a}_0^{\dag} \hat{a}_1 \big> \right) + \textrm{ h.c.} \right]  \\
&+ \left( |\psi_1(z,t)|^2 \psi_0^*(z',t) \bar{\psi}_1(z',t) + |\psi_1(z',t)|^2 \psi_0^*(z,t) \bar{\psi}_1(z,t) \right) 
\left[ e^{i\varphi} \left(\big< \hat{a}_0^{\dag} \hat{a}_1^{\dag} \hat{a}_1 \hat{a}_1 \big> - \big< \hat{a}_0^{\dag} \hat{a}_1 \big> \big< \hat{a}_1^{\dag} \hat{a}_1 \big> \right) +\textrm{ h.c.} \right] \\
&+  \psi_0^*(z,t)\bar{\psi}_1(z,t)\psi_0^*(z',t)\bar{\psi}_1(z',t)
\left[ 2 \big< \hat{a}_0^{\dag} \hat{a}_1^{\dag} \hat{a}_1 \hat{a}_0 \big> + e^{2i\varphi} \big< \hat{a}_0^{\dag} \hat{a}_0^{\dag} \hat{a}_1 \hat{a}_1 \big> + e^{-2i\varphi} \big< \hat{a}_1^{\dag} \hat{a}_1^{\dag} \hat{a}_0 \hat{a}_0 \big>
- \left( \big< e^{i\varphi} \hat{a}_0^{\dag} \hat{a}_1 + e^{-i\varphi} \hat{a}_1^{\dag} \hat{a}_0 \big> \right)^2 \right] .
\end{split}
\end{equation} 
Furthermore, we can take $\bar{\psi}_1(z,t)$ as the new $\psi_1(z,t)$, yielding $\psi_0 (z, t)
\psi_1(z, t) \in {\mathbb R}$, by changing the many-body state $\ket{\Psi}$ as follows
\begin{equation} \label{staterotation}
\ket{\Psi}=\sum_{l=0}^{N} C_l \ket{N-l,l} \rightarrow \ket{\Psi_{\varphi}}=\sum_{l=0}^{N} C_l e^{il\varphi} \ket{N-l,l}.
\end{equation}
This rotation of the state in \eqref{staterotation} corresponds to a rotation of the phase $\phi_{\beta}$ of the approximate coherent state $\ket{\beta}$, 
\begin{equation}
\ket{\beta} \rightarrow \ket{\beta e^{i\varphi}}.
\end{equation}
For a general twofold fragmented state we then get
\begin{equation}
\frac{1}{\sqrt{1+r^2+2r\cos\theta e^{-2|\beta|^2}}} \left( \ket{\beta}+re^{i\theta}\ket{-\beta} \right)
 \rightarrow \frac{1}{\sqrt{1+r^2+2r\cos\theta e^{-2|\beta|^2}}} \left( \ket{\beta e^{i\varphi}}+re^{i\theta}\ket{-\beta e^{i\varphi}} \right).
\end{equation}
Since $\phi_{\beta}=\pi/2$ before TOF, it is rotated to $\phi_{\beta}=0$ after TOF.
We can therefore summarize as follows: TOF evolution is equivalent to $\varphi=3\pi/2=-\pi/2$ rotation of state defined in \eqref{staterotation}, and we can calculate $\Delta\rho_2(z,z')$ from \eqref{densityfluctuationvarphi} for both before TOF ($\varphi=0$ or $\phi_{\beta}=\pi/2$) and after TOF ($\varphi=-\pi/2$ or $\phi_{\beta}=0$).

\subsection{Validity of the $\ket{\Psi}=\mathcal{N}(\ket{\beta}+r^{i\theta}\ket{-\beta})$ Ansatz}

As a main ingredient of our analysis, we propose an ansatz,  
$\ket{\Psi}=\mathcal{N}(\ket{\beta}+r^{i\theta}\ket{-\beta})$, to describe a twofold fragmented state 
(Eq.\,(7) in the main text).
 To test the validity of the ansatz in more detail, let us first specify the $C_l$ distribution of a general twofold fragmented state.
 In \cite{Bader}, it was shown that $|C_l|$ has a Gaussian distribution of mean $l_0$ and variance $\sigma^2$ with $\sigma\sim\mathcal{O}(\sqrt{N})$ as long as the continuum limit for the 
 $C_l$ distribution holds.
Furthermore, a general twofold fragmented state can also be written as \cite{Lee}
\begin{equation}
\ket\Psi = c \left(\ket{\rm even } + u\exp[i\theta_{\rm K}] \ket{\rm odd}\right), ~|c|^2(1+|u|^2)=1,
\end{equation}
which leads to
\begin{equation} \label{C_l dist}
\begin{cases}
|C_l|= \sqrt{2}|c| \frac{1}{(2\pi\sigma^2)^{1/4}} e^{-\frac{(l-l_0)^2}{4\sigma^2}} e^{i\phi_l} \quad \textrm{for even } l\\
|C_l|= \sqrt{2}|c||u| \frac{1}{(2\pi\sigma^2)^{1/4}} e^{-\frac{(l-l_0)^2}{4\sigma^2}} e^{i\phi_l} \quad  \textrm{for odd } l
\end{cases} ,
\end{equation}
where $C_l=|C_l|e^{i\phi_l}$, and $\phi_l$ obeys $\phi_{l+2}=\phi_l+\pi,~\phi_{l+4}=\phi_l$ from 
the condition $\textrm{sgn}(C_l C_{l+2})=-1$ \cite{Bader}. We therefore have 
\begin{equation}
\begin{cases}
\phi_{l+1}=\phi_l+\theta_k \quad \textrm{for even } l\\
\phi_{l+1}=\phi_l+\pi-\theta_k \quad  \textrm{for odd } l
\end{cases}.
\end{equation}
 We now examine to which extent using $\ket{\Psi}=\mathcal{N}(\ket{\beta}+r^{i\theta}\ket{-\beta})$ is 
 accurate for the calculation of the density-density correlations $\Delta\rho_2(z,z')$ by calculating \eqref{densityfluctuationvarphi} with \eqref{C_l dist} in terms of $\sigma,l_0$ and $c,u$.
We then compare the corresponding result for the density-density correlations obtained by using the exact large $N$ fragmented state with what we derive using Eq. (11) and the quadrature fluctuations of Eq.\,(9) in the main text. 
As we will show, we obtain two results whose difference depends on the value of the Gaussian width $\sigma$. 

  To calculate \eqref{densityfluctuationvarphi}, we perform the following approximations. When the Gaussian $|C_l|$ distribution is well localized in the interval $[0,N]$
  and $N$ is large enough to approximate $C_l$ to be continuous, then it is permissible to approximate the expectation value of $f(N-l,l)$, which is a polynomial of $\sqrt{N-l},\sqrt{l}$, as
\begin{equation} \label{cont-integ}
\begin{split}
\sum_l f(N-l,l) |C_l|^2 =& 2|c|^2 \sum_{l=0,2,\cdots} f(N-l,l) \frac{1}{\sqrt{2\pi\sigma^2}} e^{-\frac{(l-l_0)^2}{2\sigma^2}} +  2|c|^2|u|^2 \sum_{l=1,3,\cdots} f(N-l,l) \frac{1}{\sqrt{2\pi\sigma^2}} e^{-\frac{(l-l_0)^2}{2\sigma^2}} \\
 \simeq & \int^{\infty}_{-\infty} f(N-l,l) \frac{1}{\sqrt{2\pi\sigma^2}} e^{-\frac{(l-l_0)^2}{2\sigma^2}} \, dl.
\end{split}
\end{equation} 
The quantities $\big<\hat{a}_0^{\dag}\hat{a}_0\big>$,$\big<\hat{a}_1^{\dag}\hat{a}_1\big>$,$\big<\hat{a}_0^{\dag}\hat{a}_0^{\dag}\hat{a}_0\hat{a}_0\big>$,$\big<\hat{a}_1^{\dag}\hat{a}_1^{\dag}\hat{a}_1\hat{a}_1\big>$ and $\big<\hat{a}_0^{\dag}\hat{a}_1^{\dag}\hat{a}_1\hat{a}_0\big>$ 
are directly determined from the above expression. We have
\begin{equation}
\big<\hat{a}_0^{\dag}\hat{a}_0^{\dag}\hat{a}_0\hat{a}_0\big>=(N-l_0)^2-(N-l_0)+\sigma^2,~
\big<\hat{a}_1^{\dag}\hat{a}_1^{\dag}\hat{a}_1\hat{a}_1\big>=l_0^2-l_0+\sigma^2,~
\big<\hat{a}_0^{\dag}\hat{a}_1^{\dag}\hat{a}_1\hat{a}_0\big>=(N-l_0)l_0-\sigma^2
\end{equation}
with $\big<\hat{a}_0^{\dag}\hat{a}_0\big>=N-l_0$,$\big<\hat{a}_1^{\dag}\hat{a}_1\big>=l_0$.

The quantities  $\big<\hat{a}_0^{\dag}\hat{a}_1\big>$,$\big<\hat{a}_0^{\dag}\hat{a}_0^{\dag}\hat{a}_0\hat{a}_1\big>$ and $\big<\hat{a}_0^{\dag}\hat{a}_1^{\dag}\hat{a}_1\hat{a}_1\big>$, in turn,  are related to the sum 
\begin{equation}
\sum_l f(N-l,l) C_l^*C_{l+1} \simeq 2|c|^2|u|\sum_l f(N-l,l) |C_l|^2 e^{i(\phi_{l+1}-\phi_l)},  
\end{equation}
the last relation holding when 
$|C_l|\simeq|C_{l+1}|$. 
Under the proviso that $\phi_{l+4}=\phi_l$ and $|C_l|$ slowly vary so that 
$|C_l|\simeq|C_{l+4}|$, we can carry out the summation over $e^{i(\phi_{l+1}-\phi_l)} $ 
separately from the $l$ summation as
\begin{equation}
\sum_l f(N-l,l) |C_l|^2 e^{i(\phi_{l+1}-\phi_l)} \simeq \frac{1}{4}\sum_{l=0}^{3}e^{i(\phi_{l+1}-\phi_l)} 2|c|^2|u| \sum_l f(N-l,l) |C_l|^2.
\end{equation}
Therefore we have 
\begin{equation}
\sum_l f(N-l,l) C_l^*C_{l+1} \simeq \frac{1}{4}\sum_{l=0}^{3}e^{i(\phi_{l+1}-\phi_l)} 2|c|^2|u| \int^{\infty}_{-\infty} f(N-l,l) \frac{1}{\sqrt{2\pi\sigma^2}} e^{-\frac{(l-l_0)^2}{2\sigma^2}} \, dl.
\end{equation}
From the following string of phase values, $\phi_l= \cdots,0,\theta_k,\pi,\theta_k+\pi,0,\theta_k,\pi,\theta_k+\pi,\cdots$,  where $\phi_0=0$ is chosen, we obtain 
\begin{equation}
\frac{1}{4}\sum_{l=0}^{3}e^{i(\phi_{l+1}-\phi_l)}=e^{i\theta_k}+e^{i(\pi-\theta_k)}+e^{i\theta_k}+e^{-i(\pi+\theta_k)}=i\sin\theta_k .
\end{equation}
The function $f(N-l,l)$ now  
includes square roots of $N-l$ or $l$. It is therefore nontrivial to write down expectation values 
in terms of $N,l_0,\sigma$ in closed form,  
with the exception of $\big<\hat{a}_0^{\dag}\hat{a}_1\big> = 2|c|^2 u i\sqrt{(N-l_0)l_0}\sin\theta_k$. Now, $\big<\hat{a}_0^{\dag}\hat{a}_0^{\dag}\hat{a}_0\hat{a}_1\big>$ and $\big<\hat{a}_0^{\dag}\hat{a}_1^{\dag}\hat{a}_1\hat{a}_1\big>$ read
\begin{equation} \label{a0001-1}
\begin{split}
\big<\hat{a}_0^{\dag}\hat{a}_0^{\dag}\hat{a}_0\hat{a}_1\big> &\simeq 2|c|^2 u i \sin\theta_k  \int^{\infty}_{-\infty} (N-l-\frac{1}{2})\sqrt{(N-l+\frac{1}{2})(l+\frac{1}{2})} \frac{1}{\sqrt{2\pi\sigma^2}} e^{-\frac{(l-l_0)^2}{2\sigma^2}} \, dl , \\
\big<\hat{a}_0^{\dag}\hat{a}_1^{\dag}\hat{a}_1\hat{a}_1\big> &\simeq 2|c|^2 u i \sin\theta_k  \int^{\infty}_{-\infty} (l-\frac{1}{2})\sqrt{(N-l+\frac{1}{2})(l+\frac{1}{2})} \frac{1}{\sqrt{2\pi\sigma^2}} e^{-\frac{(l-l_0)^2}{2\sigma^2}} \, dl.
\end{split}
\end{equation}
From the following integrals 
\begin{equation}
\int^{\infty}_{-\infty} l \frac{1}{\sqrt{2\pi\sigma^2}} e^{-\frac{(l-l_0)^2}{2\sigma^2}} \, dl = l_0 , \quad
\int^{\infty}_{-\infty} l^2 \frac{1}{\sqrt{2\pi\sigma^2}} e^{-\frac{(l-l_0)^2}{2\sigma^2}} \, dl = l_0^2 + \sigma^2, 
\end{equation}
 we can power expand  $\label{a0001_1}$ as
\begin{equation} \label{a0001-2}
\begin{split}
\big<\hat{a}_0^{\dag}\hat{a}_0^{\dag}\hat{a}_0\hat{a}_1\big> &\simeq 2|c|^2 u i \sin\theta_k\left[ (N-l_0)\sqrt{(N-l_0)l_0}  + \mathcal{O}(N-l_0) +\mathcal{O}(\sigma^2) \right] , \\
\big<\hat{a}_0^{\dag}\hat{a}_1^{\dag}\hat{a}_1\hat{a}_1\big> &\simeq 2|c|^2 u i \sin\theta_k\left[ l_0\sqrt{(N-l_0)l_0}  + \mathcal{O}(N-l_0) +\mathcal{O}(\sigma^2) \right] .
\end{split}
\end{equation}
The first, highest order, terms are matching $\big<\hat{a}_0^{\dag}\hat{a}_0\big>\big<\hat{a}_0^{\dag}\hat{a}_1\big>\propto (N-l_0)\sqrt{(N-l_0)l_0}$ and $\big<\hat{a}_0^{\dag}\hat{a}_1\big>\big<\hat{a}_1^{\dag}\hat{a}_1\big>\propto l_0\sqrt{(N-l_0)l_0}$; 
we assume $l_0<N/2$.
Finally, for $\big<\hat{a}_0^{\dag}\hat{a}_0^{\dag}\hat{a}_1\hat{a}_1\big>$ ($\big<\hat{a}_1^{\dag}\hat{a}_1^{\dag}\hat{a}_0\hat{a}_0\big>$ is given by the complex conjugate), with slowly varying $|C_l|$, 
we have
\begin{equation}
\big<\hat{a}_0^{\dag}\hat{a}_0^{\dag}\hat{a}_1\hat{a}_1\big> \simeq \frac{1}{4}\sum_{l=0}^{3}e^{i(\phi_{l+2}-\phi_l)} \int^{\infty}_{-\infty} \sqrt{(N-l)(N-l+1)l(l+1)} \frac{1}{\sqrt{2\pi\sigma^2}} e^{-\frac{(l-l_0)^2}{2\sigma^2}} \, dl.
\end{equation}
A expansion similar to that leading from \eqref{a0001-1} to \eqref{a0001-2}, and $\frac{1}{4}\sum_{l=0}^{3}e^{i(\phi_{l+2}-\phi_l)}=-1$, yields 
\begin{equation}
\big<\hat{a}_0^{\dag}\hat{a}_0^{\dag}\hat{a}_1\hat{a}_1\big>\simeq -(N-l_0)l_0  + \mathcal{O}(N-l_0) +\mathcal{O}(\sigma^2).
\end{equation}
Summing up, Eq.\,\eqref{densityfluctuationvarphi} becomes
\begin{equation} \label{SMresult1}
\begin{split}
& \Delta\rho_2(z,z',t) = |\psi_0(z,t)|^2|\psi_0(z',t)|^2 \left( \sigma^2-(N-l_0) \right)
 + |\psi_1(z,t)|^2|\psi_1(z',t)|^2 \left( \sigma^2-l_0 \right) \\
&+ \left( |\psi_0(z,t)|^2|\psi_1(z',t)|^2 + |\psi_1(z,t)|^2|\psi_0(z',t)|^2 \right) \sigma^2 \\
&+ 2|c|^2 u i \sin\theta_k \left( |\psi_0(z,t)|^2 \psi_0^*(z',t) \bar{\psi}_1(z',t)  + |\psi_0(z',t)|^2 \psi_0^*(z,t) \bar{\psi}_1(z,t)  \right)
\left[\mathcal{O}(N-l_0) +\mathcal{O}(\sigma^2)\right]   \\
&+ 2|c|^2 u i \sin\theta_k \left( |\psi_1(z,t)|^2 \psi_0^*(z',t) \bar{\psi}_1(z',t) + |\psi_1(z',t)|^2 \psi_0^*(z,t) \bar{\psi}_1(z,t) \right) 
\left[\mathcal{O}(N-l_0) +\mathcal{O}(\sigma^2)\right] \\
&+ 4\psi_0^*(z,t)\bar{\psi}_1(z,t)\psi_0^*(z',t)\bar{\psi}_1(z',t)\left[ (N-l_0)l_0\sin^2\varphi\left(1-(2|c|^2 u\sin\theta_k)^2\right) + \mathcal{O}(N-l_0) +\mathcal{O}(\sigma^2) \right].
\end{split}
\end{equation} 
Provided the orbitals $\psi_0(z,t)$ and $\psi_1(z,t)$ have large overlap, as to be expected 
in a single trap, the following four terms have similar order of magnitude, 
\begin{equation} \label{simplifycondition2}
\psi_0^*(z,t)\bar{\psi}_1(z,t)\psi_0^*(z',t)\bar{\psi}_1(z',t) ~ \sim ~ |\psi_0(z,t)|^2|\psi_0(z',t)|^2, ~|\psi_1(z,t)|^2|\psi_1(z',t)|^2, ~|\psi_0(z,t)|^2|\psi_1(z',t)|^2.
\end{equation}
This leads to
\begin{equation} \label{SMresult2}
\Delta\rho_2(z,z',t) = 4\psi_0^*(z,t)\bar{\psi}_1(z,t)\psi_0^*(z',t)\bar{\psi}_1(z',t)(N-l_0)l_0\sin^2\varphi\left(1-(2|c|^2 u\sin\theta_k)^2\right) + \mathcal{O}(N-l_0) +\mathcal{O}(\sigma^2).
\end{equation}
Therefore, since the first term of \eqref{SMresult2} is proportional to $(N-l_0)l_0$, and because 
we assume 
$(N-l_0)l_0\gg N-l_0,~(N-l_0)l_0\gg \sigma^2$,
we obtain $\Delta\rho_2(z,z')$ to lowest order as
\begin{equation} \label{simple}
\Delta\rho_2(z,z') \simeq  4\psi_0^*(z,t)\bar{\psi}_1(z,t)\psi_0^*(z',t)\bar{\psi}_1(z',t)(N-l_0)l_0\sin^2\varphi\left(1-(2|c|^2 u\sin\theta_k)^2\right) .
\end{equation}
To compare with the result in Eq.\,(11) of the main text, we employ the relation between 
$r,\theta$ and $u,\theta_k$ specified for large $|\beta|$ by Eq.\,(8). 
Then $2|c|^2 u\sin\theta_k=2u/(1+u^2)\sin\theta_k$ can be expressed in terms of $u,\theta_k$ as
\begin{equation}
\frac{1-r^2}{1+r^2}=\frac{2u\lambda_{\beta}}{1+u^2\lambda_{\beta}^2}\sin\theta_k=\left(\frac{2u}{1+u^2}\sin\theta_k\right)\frac{1+u^2}{\lambda_{\beta}(\lambda_{\beta}^{-2}+u^2)}.
\end{equation}
In the  $\lambda_{\beta}=\sqrt{(1+e^{-2|\beta|^2})/(1-e^{-2|\beta|^2})}\rightarrow 1$ limit 
(equivalent to $e^{-2|\beta|^2}\rightarrow 0$ and small overlap of $\ket{\beta}$ and $\ket{-\beta}$), 
this becomes
\begin{equation}
\frac{1-r^2}{1+r^2}=\frac{2u}{1+u^2}\sin\theta_k=2|c|^2 u \sin\theta_k.
\end{equation}
The result does not depend on $\theta$. Thus we see that when the overlap between $\ket{\beta}$ and $\ket{-\beta}$ gets smaller, the effect of $\theta$ on $\Delta\rho_2(z,z')$ becomes negligible.
 In this limit \eqref{simple} in terms of $r$, with $|\beta|^2=l_0$, is given as
\begin{equation} \label{simple2}
\Delta\rho_2(z,z') \simeq  4\psi_0^*(z,t)\bar{\psi}_1(z,t)\psi_0^*(z',t)\bar{\psi}_1(z',t)(N-l_0)l_0\sin^2\varphi \frac{4r^2}{(1+r^2)^2},
\end{equation}
which agrees with the $r=1$ result of the main text in Eq.\,(11) 
as long as $(N-l_0)l_0 \gg N-l_0, \sigma^2$. 

We conclude that $(N-l_0)l_0 \gg N-l_0, \sigma^2$, and the large single-trap overlap between 
the orbitals $\psi_0(z,t)$ and $\psi_1(z,t)$ are the conditions to utilize the representation 
$\ket{\Psi}=\mathcal{N}(\ket{\beta}+r^{i\theta}\ket{-\beta})$ for a given twofold fragmented state.

\end{widetext}


\begin{thebibliography}{9999}
\bibitem{note} 
Nonclassical superposition states of appreciable size 
can potentially exist even for very large $N$. 
According to \cite{Chatterjee} ${\mathcal F}\propto 1/\sqrt N$ 
and hence, we would have $|\beta|^2 \propto \sqrt N$. The many-body states, for these small $\mathcal F$,  can however no longer accurately be described as phase-space or ladder-coherent state superpositions, and require a separate analysis.
\bibitem{squeezing} Note that TOF can equivalently be viewed as a conversion between ``position" and ``momentum" squeezing. 
\end{thebibliography}
\end{document}